\documentclass{elsart}
\usepackage{epsfig}
\usepackage{graphicx} 
\usepackage{dcolumn}  
\graphicspath{{ps}}

\def\mySpecialText{DRAFT version 2.4 2004/June/24}

\def\myspecial#1{}                   

\def\calL{{\mathcal L}}

\def\Mbc{M_{\rm bc}}
\def\LR{{\mathcal R}}

\begin{document}

\begin{frontmatter}
\epsfysize3cm

\hspace{-9.5cm}
\epsfbox{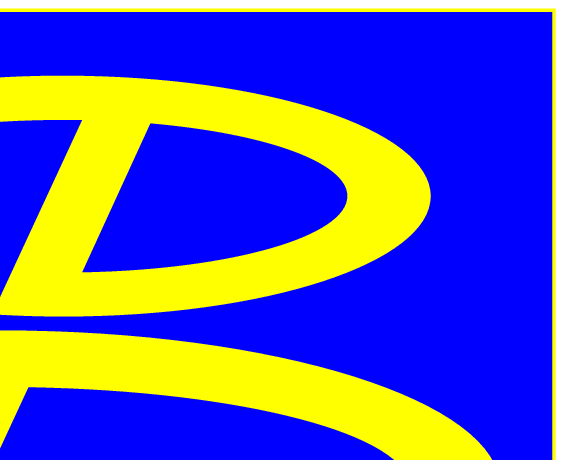}    

\vskip -2cm
\noindent
\hspace*{9.5cm}KEK Preprint 2004-25 \\
\hspace*{9.5cm}Belle Preprint 2004-18 \\

\vskip 1.3cm
\myspecial{!userdict begin /bop-hook{gsave 300 50 translate 5 rotate
    /Times-Roman findfont 18 scalefont setfont
    0 0 moveto 0.70 setgray
    (\mySpecialText)
    show grestore}def end}

\title{\quad\\[0.5cm] \Large
Observation of the Decays $B^0\to K^+\pi^-\pi^0$ and $B^0\to \rho^- K^+$. 
}

\collab{Belle Collaboration}
  \author[Taiwan]{P.~Chang}, 
  \author[KEK]{K.~Abe}, 
  \author[TohokuGakuin]{K.~Abe}, 
  \author[KEK]{T.~Abe}, 
  \author[Tokyo]{H.~Aihara}, 
  \author[Tsukuba]{Y.~Asano}, 
  \author[BINP]{V.~Aulchenko}, 
  \author[ITEP]{T.~Aushev}, 
  \author[Tata]{T.~Aziz}, 
  \author[Cincinnati]{S.~Bahinipati}, 
% ...

\address[BINP]{Budker Institute of Nuclear Physics, Novosibirsk, Russia}
% ...
\address[Cincinnati]{University of Cincinnati, Cincinnati, OH, USA}
% ...
\address[KEK]{High Energy Accelerator Research Organization (KEK), Tsukuba, Japan}
% ...
\address[ITEP]{Institute for Theoretical and Experimental Physics, Moscow, Russia}
% ...
\address[Taiwan]{Department of Physics, National Taiwan University, Taipei, Taiwan}
% ...
\address[Tata]{Tata Institute of Fundamental Research, Bombay, India}
% ...
\address[TohokuGakuin]{Tohoku Gakuin University, Tagajo, Japan}
% ...
\address[Tokyo]{Department of Physics, University of Tokyo, Tokyo, Japan}
% ...
\address[Tsukuba]{University of Tsukuba, Tsukuba, Japan}
% ...
\thanks[Fermilab]{on leave from Fermi National Accelerator Laboratory, Batavia, IL, USA}
\thanks[NovaGorica]{on leave from Nova Gorica Polytechnic, Nova Gorica, Slovenia}

\normalsize

\begin{abstract}
We report the observation of $B^0$ decays to the $K^+\pi^-\pi^0$ final
 state using a data sample of
 78 fb$^{-1}$  collected by the Belle detector at the KEKB 
$e^+e^-$ collider. With no assumptions about intermediate states in
 the decay, the 
branching fraction is measured to be $(36.6^{+4.2}_{-4.3}\pm 3.0)\times 
10^{-6}$. 
We also search for $B$ decays to intermediate two-body states with the same  $K^+\pi^-\pi^0$ final state.
Significant $B$ signals are observed in the $\rho(770)^- K^+$ and 
$K^*(892)^+\pi^-$ channels, with  branching fractions of
$(15.1^{+3.4+1.4+2.0}_{-3.3-1.5-2.1})\times 10^{-6}$ and 
$(14.8^{+4.6+1.5+2.4}_{-4.4-1.0-0.9})\times 10^{-6}$, respectively.
The first  error is statistical, the second is systematic and the third 
 is due to the largest possible  interference.
Contributions from other possible two-body states will be discussed. 
No  $CP$ asymmetry is found
in the inclusive $K^+\pi^-\pi^0$ or $\rho^-K^+$ modes, and we set 90\%
confidence level
bounds on the asymmetry  of $-0.12<A_{CP}<0.26$ and
$-0.18<A_{CP}<0.64$, respectively.

\vspace{3\parskip}
\end{abstract}


\end{frontmatter}


\section{Introduction}

Recently, observations of large branching fractions for three-body charmless hadronic decays of $B$ 
mesons have been reported by the $B$ factory experiments 
\cite{khh,kshh,ppk,bkhh,ckstp,brhok}.
In the mesonic decays $B\to K\pi^+\pi^-$ and $B\to K K^+K^-$, a large 
fraction of the decays proceed through intermediate two-body 
decay processes, such as $B^+\to K^*(892)^0\pi^+$, $K^*(892)^0\to K^+\pi^-$ and 
$B^+\to \phi K^+, \phi \to K^+ K^-$. However,  higher mass $K^+\pi^-$,
$\pi^+\pi^-$ and $K^+K^-$ states  may contribute but are not  clearly
identified due to limited 
statistics. Moreover, the broad $K^+K^-$ mass spectrum above 1.5 GeV/$c^2$ in
$B^+\to K^+K^+K^-$ 
 suggests a large non-resonant $B^+\to K^+K^+K^-$ contribution. In the baryonic
decay $B^+\to p\bar{p}K^+$, a simple phase-space model fails to describe the $p\bar{p}$
mass spectrum, which may be explained by a baryonic form factor model 
\cite{hou} or by an additional, unknown resonance around 2 GeV/$c^2$.      
  These studies of three-body decays have provided new information on the 
mechanism of B meson decay, and 
provide opportunities  to search for unknown $B$ meson decays and to understand 
the interference between them. In this paper we report on a study
of $B$ meson decays to $K^+\pi^-\pi^0$, independently of possible  
  intermediate states.
In addition, we also present  results of a search for  quasi-two-body intermediate states.
Inclusion of charge conjugate modes is always implied in this letter unless 
otherwise specified. 
The results are obtained from data collected by
the Belle detector \cite{belle} at 
the KEKB asymmetric $e^+e^-$ storage ring \cite{accel}. 
The data sample corresponds to an integrated luminosity of
78 fb$^{-1}$ and contains 85.0 million $B\overline{B}$ pairs
at the $\Upsilon(4S)$ resonance. 

\section{Apparatus and Event Selection}

The Belle detector is a large-solid-angle general purpose spectrometer based on
a 1.5 T superconducting solenoidal magnet. 
Charged tracks are reconstructed 
with a three layer double-sided silicon vertex detector (SVD)
and a central drift chamber (CDC) that consists of 
50 layers segmented into 6 axial and 5 stereo superlayers. 
The CDC covers the polar angle range between $17^\circ$ and 
$150^\circ$ in the laboratory frame and, together with the SVD, 
gives a transverse momentum resolution of 
$(\sigma_{p_t}/p_t)^2 = (0.0019 \,p_t)^2 +(0.0030)^2,$  
where $p_t$ and $\sigma_{p_t}$ are in GeV/$c$. 
Charged hadron  identification is performed using
a combination of three devices: 
an array  of 1188 aerogel \v{C}erenkov counters (ACC) 
covering the momentum range 1--4 GeV/$c$, 
a time-of-flight scintillation counter system (TOF) 
for track momenta below 1.5 GeV/$c$, 
and $dE/dx$ information from the CDC 
for particles with low or high momenta. 
Situated between these devices and the solenoid coil is an
electromagnetic calorimeter (ECL) consisting of 8736 CsI(T$\ell$) crystals
with a typical front-surface cross-section of 
$5.5 \times 5.5 $ cm$^2$ and 
a depth of $16.2\,X_0$. 
The ECL provides a photon energy resolution of 
$(\sigma_E/E)^2 = 0.013^2 + (0.0007/E)^2 + (0.008/E^{1/4})^2$, where $E$ 
and $\sigma_E$ are in GeV.  An instrumented iron flux return outside the 
solenoid coil is used for muon and $K_L$ detection.  
A detailed description of the Belle detector 
can be found in Ref. \cite{belle}. 

Charged tracks are required to come from the collision point and have
transverse momenta, $p_t$, above 100 MeV/$c$. 
The accepted tracks are then refitted with their vertex position
constrained to the run-averaged profile of $B$ meson 
decay vertices in the transverse plane. 
Charged $K$ and $\pi$ mesons
are identified by combining information from the CDC ($dE/dx$),
the TOF and the ACC to form a $K(\pi)$ likelihood $L_K(L_\pi)$. 
Discrimination between kaons and pions is
achieved through 
the likelihood ratio $L_{K}$/($L_{\pi}+L_{K}$).
The performance of the charged hadron identification is studied using 
a kinematically selected high momentum $D^{*+}$ data sample, where
$D^{*+}\to D^0\pi^+$, $D^0\to K^-\pi^+$. 
We measure the pion and kaon identification efficiencies and their 
fake rates as functions of track momentum.  The typical kaon and pion
identification efficiencies for 1 GeV/$c$ tracks are 
$(87.9\pm 0.6)$\% and $(89.4\pm 0.6)$\%, respectively.
The rate for true pions to be misidentified as kaons is $(9.0 \pm 0.5)$\%, 
while the rate for true kaons to be misidentified as pions is
$(10.0\pm 0.6)$\%. Charged tracks which are positively identified as electrons
and muons are rejected. Candidate neutral pions are selected by requiring the
two-photon invariant mass to be in the mass window 
$0.118$ GeV/$c^2<M(\gamma\gamma)<0.150$ GeV/$c^2$, corresponding to  
$\pm 2.5 \sigma$ mass resolution  with momentum above 2 GeV/$c$. 
The momentum of each photon is then readjusted,
constraining the mass of the photon pair to be the nominal $\pi^0$ mass. To 
reduce the background from soft photons, each photon is required to have 
energy above 50 MeV and the minimum $\pi^0$ momentum is 200 MeV/$c$.

Candidate $B$ mesons are identified using the beam constrained mass,
$\Mbc =  \sqrt{E^2_{\mbox{\scriptsize beam}} - P_B^2}$,
and the energy difference, $\Delta E = E_B  - E_{\mbox{\scriptsize beam}}$, 
where $E_{\mbox{\scriptsize beam}}$ is run-dependent and  determined  from 
$B\to D^{(*)}\pi$ events, 
and $P_{B}$ and $E_B$ are the momentum and energy of  the
$B$ candidate in the $\Upsilon(4S)$ rest frame. 
The parameterizations of the signal in $\Mbc$ and 
$\Delta E$ are determined by a GEAN-based Monte Carlo (MC) \cite{geant} 
simulation of non-resonant $B^0\to K^+\pi^-\pi^0$ decays, and various quasi-two-body
decays to the $K^+\pi^-\pi^0$ final state. 
The signal parameterization is verified using the data and MC samples of
$B^+\to \overline{D}{}^0\pi^+, \overline{D}{}^0\to K^+\pi^-\pi^0$
candidates. Our MC overestimates the $\Mbc$ resolution by 8\% but 
underestimates the $\Delta E$ resolution by 9\% to 15\%, depending on 
the kinematics of the $K^+\pi^-\pi^0$ events. The MC based signal probability
density functions (PDF) are
 readjusted accordingly.   
  
The Gaussian width of the signal in $\Mbc$ is  about 3.0 MeV/$c^2$, 
which is primarily due to  the beam energy spread.  
The $\Delta E$ distribution is found to be  asymmetric,
with a  tail on the lower side due to 
$\gamma$ interactions with material in front of the calorimeter, and
shower leakage out of the back side of the crystals. As a result, 
the $\Delta E$ 
resolution and the tail distribution strongly depend  on the $\pi^0$ energy;
 the $\Delta E$ width ranges from  20 MeV to 33 MeV. 
In the inclusive $K^+\pi^-\pi^0$ study, since  the $\pi^0$ energy
distribution for the signal is not known a priori, the  data is 
divided into
three samples: $P(\pi^0)<0.5$ GeV/$c$, $0.5$ GeV/$c<P(\pi^0)<1.5$ GeV/$c$ and
$P(\pi^0)>1.5$ GeV/$c$. The $\Delta E$ distribution in each sample is modeled 
with a Crystal Ball lineshape \cite{crystal} with parameters determined from  
MC.
Events with $\Mbc>5.2$ GeV/$c^2$ and
$|\Delta E| < 0.3$ GeV are selected for the final analysis. 
The signal region is defined as $\Mbc >5.27$ GeV/$c^2$ and 
$-0.10$ GeV $<\Delta E<0.08$ GeV. 
Events located in the region $\Mbc <5.265$  GeV/$c^2$
are defined as sideband events and are used for background studies.
When more than
one $B^0$ candidate is found in an event, the candidate having
the smallest sum of the $\chi^2$ from the vertex fit and $\pi^0$ mass 
constrained fit is selected. 

The dominant background for three-body $B$ decay events 
comes from the $e^+e^-\rightarrow q\bar{q}$ continuum, where $q= u, d, s$ or 
$c$. 
In order to reduce this background, several shape variables are chosen to
distinguish spherical $B\overline{B}$ events from jet-like continuum events. 
Five modified Fox-Wolfram moments \cite{sfw} and a measure of the momentum 
transverse to the event thrust axis  
($S_\perp$) \cite{sperp}
are combined into a Fisher discriminant.  The PDFs 
 for this discriminant and $\cos\theta_B$, where $\theta_B$ is the angle 
between the $B$ flight direction and the beam direction 
in the $\Upsilon(4S)$ rest frame,  are 
obtained using events in the signal
and sideband regions from MC simulations for signal and $q \bar{q}$ 
background. These two variables are then combined to form 
a likelihood ratio $\LR = {\calL}_s/({\calL}_s + {\calL}_{q \bar{q}})$, 
where ${\calL}_{s (q \bar{q})}$ is 
the product of signal ($q \bar{q}$) probability densities. 
Continuum background is suppressed by requiring $\LR>0.9$, based on a study
of the signal significance ($N_S/\sqrt{N_S+N_B}$) using a   
MC sample, where $N_S$ and $N_B$ are signal 
and background yields, respectively. This $\LR$ requirement retains
45\% of the signal and removes 97\% of the continuum events. 
The effect of the $\LR$ cut is studied by 
comparing $B^+\to \overline{D}{}^0\pi^+$
in data and MC, for different values of $\LR$. A systematic error of 3\%
is obtained for the $\LR$ cut. 

\begin{figure}[th]
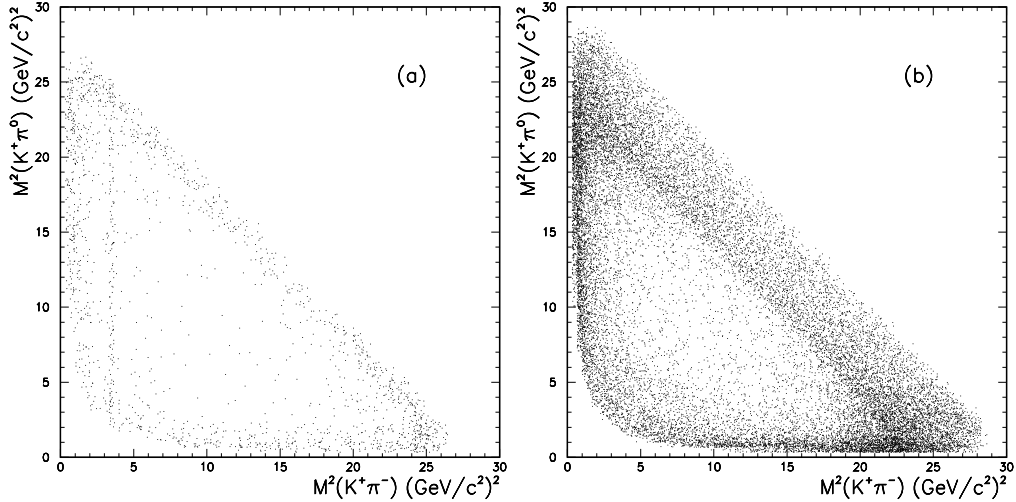

\begin{center}
\epsfig{file=dal1.epsi,width=2.6in,height=2.6in}
\epsfig{file=dal2.epsi,width=2.6in,height=2.6in}
\caption{Dalitz plot of $B^0\to K^+\pi^-\pi^0$ candidates from (a) 
the $B$ signal 
region  and (b) the $\Mbc$ sideband region. The enhancement around 
$M^2(K^+\pi^-) = 3.5 $ ({\rm GeV}/$c^2)^2$ is from
$B^0\to \overline{D}{}^0\pi^0$ events. 
 }
\label{fig:dalitz}
\end{center}
\end{figure}

Figure \ref{fig:dalitz} shows the Dalitz plot distribution of $K^+\pi^-\pi^0$ candidates
in the $\Delta E-\Mbc$ signal region and $\Mbc$ sideband region.
$K^+\pi^-\pi^0$  candidates  populate 
 the three edges of the Dalitz plot, indicating the existence of quasi-two-body
 intermediate states. Moreover, there is an enhancement near $M^2(K^+\pi^-) = 
3.5$ (GeV/$c^2)^2$ in Fig. \ref{fig:dalitz}(a), which is due to the decay  $B^0\to \overline{D}{}^0\pi^0, 
\overline{D}{}^0 \to K^+\pi^-$. To restrict the study to charmless $B$ decays,
events with a $K^+\pi^-$ mass within 50 MeV/$c^2$ of the nominal 
$\overline{D}{}^0$ mass are rejected.     

\section{Inclusive $K^+\pi^-\pi^0$ Yield} 

The final signal yields are obtained  from fits to the $\Delta E$ and 
$\Mbc$ distributions.
In addition to continuum background, the final sample contains background from
$\Upsilon (4S) \to B\overline{B}$ events.  The  $\Delta E$ and $\Mbc$ shapes of this $B\overline{B}$ 
background are modeled with 
smooth histograms, generated from a large GEANT based $B\overline{B}$ MC sample, 
 which includes $b\to c$ transitions and charmless $B$ decays.  
The continuum $\Delta E$ background shape is modeled by either
a first or second order polynomial, determined from the $\Mbc$ sideband data. 
The continuum $\Mbc$ background shape is modeled with
an ARGUS function \cite{argus} with parameters determined from events 
outside  the $\Delta E$ signal region. One-dimensional binned likelihood fits 
to $\Delta E$ and $\Mbc$  are performed 
using signal, continuum background and $B\overline{B}$ background
PDFs for events in the $\Mbc$ and $\Delta E$ signal region, 
respectively. Since the $\Mbc$ 
shapes from $B\overline{B}$ background are difficult to  distinguish from  signal 
shapes, the signal yields are estimated using the $\Delta E$ fit and cross
checked by the $\Mbc$ fit, where the $B\overline{B}$ background fraction is fixed to 
the MC expectation. 

\begin{figure}[h]
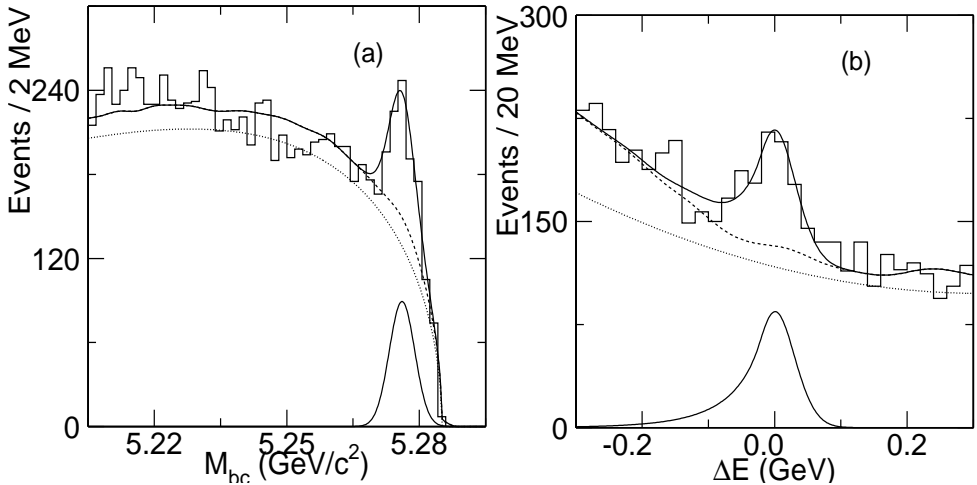

\begin{center}
\epsfig{file=mb_3b.epsi,width=2.5in,height=2.5in}
\epsfig{file=de_3b.epsi,width=2.5in,height=2.5in}
\caption{ (a) $\Mbc$  and (b) $\Delta E$   distributions for inclusive
$K^+\pi^-\pi^0$ events. Solid lines in the figures represent
the result of the fit and the signal contribution. Dashed lines show
the total background contribution, while dotted lines indicate the continuum
contribution. The $\Delta E$ lines in (b) show the sum of the fit results  
to the three subsamples. A sizeable $B^0\to K^+\pi^-$ and $B^+\to K^+\pi^0$ 
feed-down at $\Delta E > 0.2$ GeV is found in the MC. This 
contribution is included in the fit.
  }
\label{fig:inc_fit}
\end{center}
\end{figure}

Table \ref{tab:inc_re} summarizes the fit result of the inclusive 
$K^+\pi^-\pi^0$ sample  
with the statistical significance ($\Sigma$) defined as
$\sqrt{-2\ln(\calL_0/\calL_{\rm max})}$, 
where $\calL_0$ and $\calL_{\rm max}$ denote 
the likelihood values at zero  yield
and the best fit numbers, respectively. The sum of the signal yield from the
$\Delta E$  fits to the three subsamples, $386 \pm 44$, is 
consistent
with the yield from the $\Mbc$ fit, $369\pm 35$,
which has a smaller signal efficiency than the $\Delta E$ fit due to the 
tighter $\Delta E$ requirement. The corresponding projections of the fits 
are shown in Fig. \ref{fig:inc_fit}. 
Furthermore, 
a consistent result is obtained when the $B\overline{B}$ fraction is fixed 
according to the MC expectation.

\begin{table}[t]
\caption{Fit results for inclusive $K^+\pi^-\pi^0$ events. Column 3 and 4  
list the  signal yield and the statistical 
significance. The $B\overline{B}$ fraction is fixed to the MC expectation in 
the $\Mbc$ fit, while it is allowed to float in the $\Delta E$ fit. The last 
row shows  the sum of the three  $\Delta E$ fits. }
\vspace{0.1cm}
\label{tab:inc_re}
\begin{center}
\begin{tabular}{|cc|cc|} \hline
 &$\pi^0$ momentum (GeV/$c$) &   Signal Yield  &  Significance \\ \hline
$\Mbc$ fit & No Requirement & $369\pm 35$ & 11.4    \\ \hline
&$p<0.5$ & $ 31^{+13}_{-12}\pm 4$ & 2.6 \\
$\Delta E$ fit& $0.5<p<1.5$  &$ 93\pm 22 ^{+6}_{-7} $ &4.6  \\
&$p>1.5$ & $ 262\pm 36\pm 13$   &7.8  \\ 
&Sum   &$386 \pm 44^{+14}_{-15} $&9.4  \\\hline

\end{tabular}
\end{center}
\end{table}

\begin{figure}[t]
\begin{center}
\epsfig{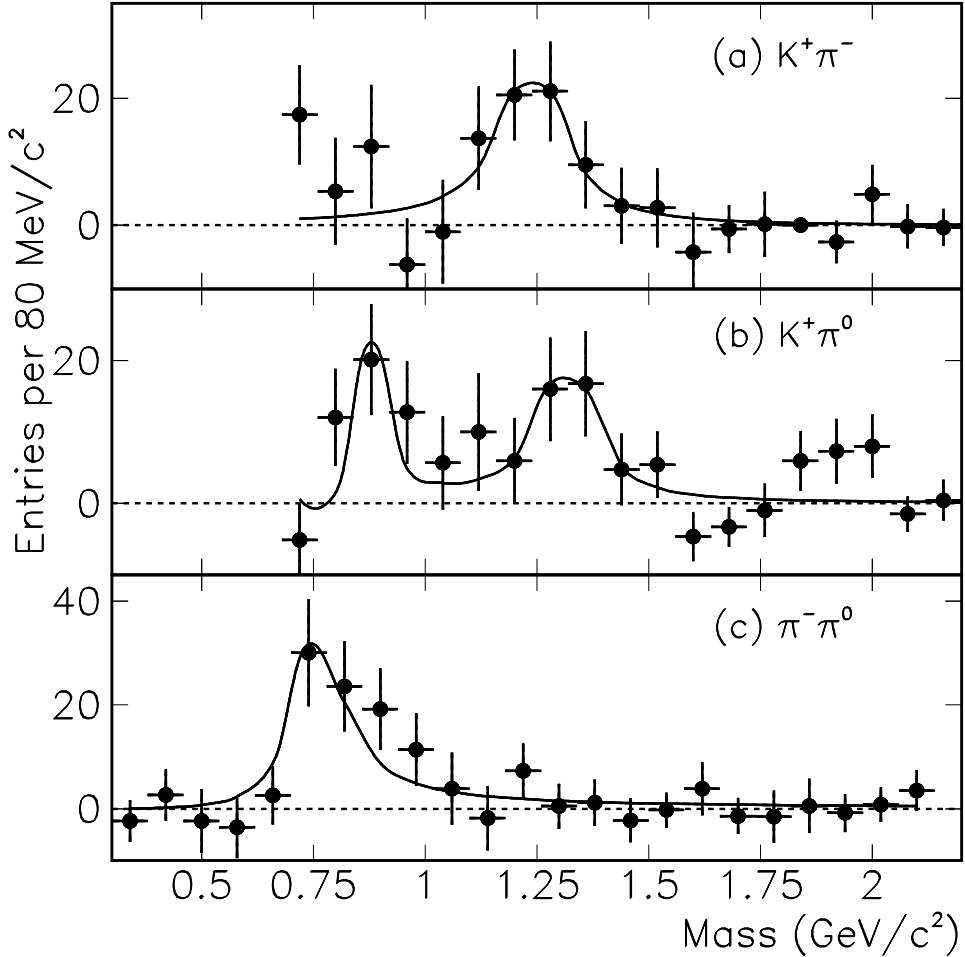}
\caption{$B$ yields from $\Delta E$ fits as a function of (a) $K^+\pi^-$, (b)
$K^+\pi^0$ and (c) $\pi^-\pi^0$.  Each two-body mass is examined after requiring
the other  two-body masses to be large                  
($>1.6$ GeV/$c^2$ for $K\pi$ and $>1.1$ GeV/$c^2$ for 
$\pi^-\pi^0$). 
Dotted points are data; the superimposed curves in (b) and
(c) are the projection curves based on $K^*(892)^+$ and $\rho(770)^-$,
The enhancements between
1.1  to 1.5 GeV/$c^2$ in (a) and (b) are modeled with  Breit-Wigner functions.
 } 
\label{fig:deind}
\end{center}
\end{figure}

\section{Two-body Intermediate States}

We perform a search for quasi-two-body decays in the $K^+ \pi^- \pi^0$ final
state, including $B$ decays to a pseudoscalar ($K$ or $\pi$) and a vector meson
($\rho(770)$ or $K^*(892)$) and other possible intermediate states with higher 
mass resonances. Three pseudoscalar-vector (PV) modes are considered:
$K^*(892)^0\pi^0$, $K^*(892)^+\pi^-,$ and $\rho(770)^- K^+$.  
Figure \ref{fig:deind} shows the $B$ signal yields from the $\Delta E$ fit as  
functions of $K^+\pi^-$, $K^+\pi^0$ and $\pi^-\pi^0$ masses.  
To eliminate  cross-talk between decay modes, each
two-body mass is examined after requiring the other
two-body masses to be large 
($>1.6$ GeV/$c^2$ for $K\pi$ and $>1.1$ GeV/$c^2$ for 
$\pi^-\pi^0$). The $\Delta E$ signal PDFs are obtained from MC simulations of 
$B^0\to K^*_0(1430)^0 \pi^0$, $B^0\to K^*_0(1430)^+\pi^-$ and 
$B^0\to \rho(770)^-K^+$ for the $K^+\pi^-$, $K^+\pi^0$ and $\pi^-\pi^0$ cases,
respectively.

 In the $K^+\pi^-$ sample, a large enhancement is observed  between 1.0  and 1.6
GeV/$c^2$, peaking around 1.2 to 1.4 GeV/$c^2$.  More structure is observed
in the $K^+\pi^0$ spectrum. An enhancement is seen in the  $K^*(892)^+$ mass 
region, in the region from 1.2 to 1.4 GeV/$c^2$
and possibly between 1.8 and 2.1 GeV/$c^2$.  
In the $\pi^-\pi^0$ 
sample, a clear excess is seen in the $\rho(770)^-$ signal region. Although the enhancement
  between 1.1  and 1.6 GeV/$c^2$ is observed in both $K^+\pi^-$ and
$K^+\pi^0$ spectra, these higher mass $K\pi$ states cannot be identified 
without performing an angular analysis that requires much more data.
Possible candidates are $K^*(1410)$, 
$K^*_0(1430)$ and $K^*_2(1430)$. 
Earlier studies of $B^+\to K^+\pi^+\pi^-$ \cite{khh} 
and $B^0\to K^0\pi^+\pi^-$ \cite{ckstp} 
decays  also observed large quasi-two-body $B$ decays with $K^*_x(K\pi)$ mesons
in
 the final state. If the enhancements observed in the  $K^+\pi^0$ mass spectrum are due to such $K^*_x$ mesons, one would expect the same resonances to appear in the 
$K^0 \pi^+$ mode. The first two enhancements seen in the $K^+ \pi^0$ mode (Fig. \ref{fig:deind}(b)) are indeed observed in the 
$B^0\to K^0\pi^+\pi^-$ 
analysis~\cite{kshh}. However, the third 
enhancement, between 1.8 and 2.1 GeV/$c^2$, does not appear in the $K^0\pi^+\pi^-$ mode.
This enhancement, which has a signal yield of $22^{+9}_{-8}$ events and 
a significance of  
$3 \sigma$,  may be either a statistical fluctuation, or originate from  
$K_2(1820)$, $K^*_4(2045)$ or from a doubly Cabibbo suppressed $D^+$  decay.
More data are needed to  clarify the current situation.     

\begin{figure}[t]
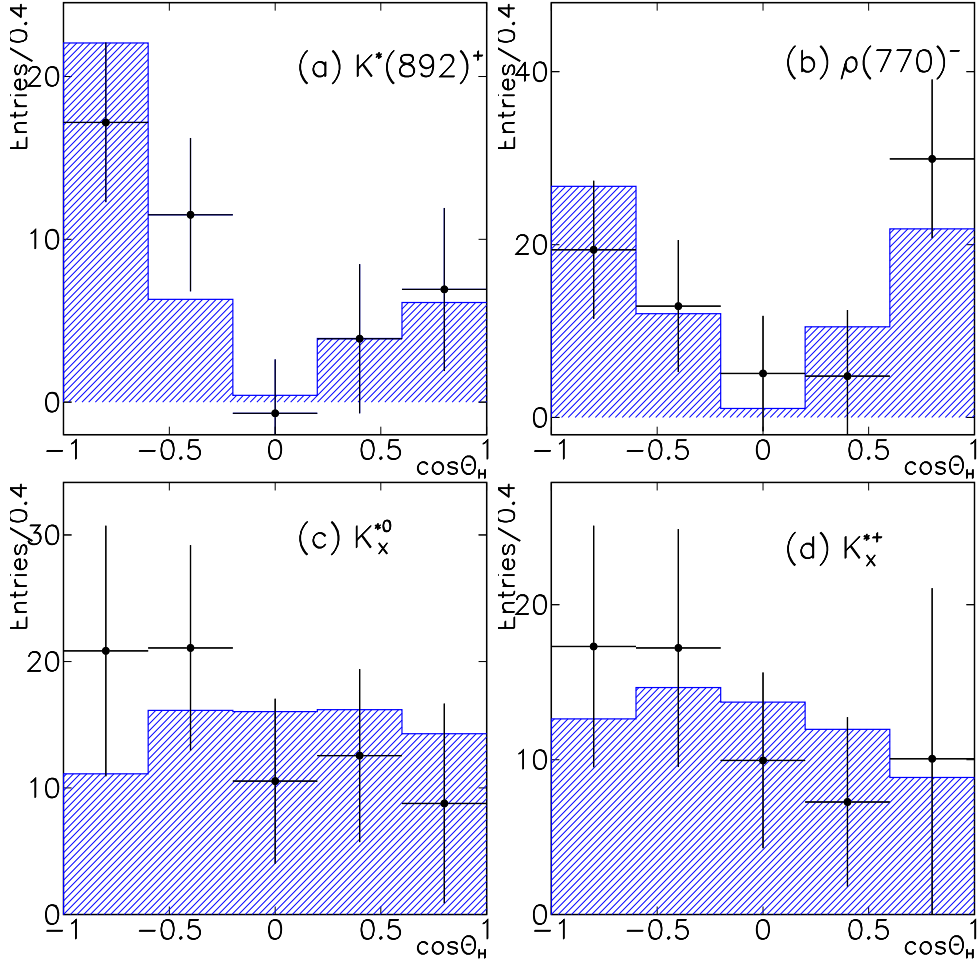

\begin{center}
\epsfig{file=he_a.epsi,width=2.5in,height=2.5in}
\epsfig{file=he_b.epsi,width=2.5in,height=2.5in}
\epsfig{file=he_c.epsi,width=2.5in,height=2.5in}
\epsfig{file=he_d.epsi,width=2.5in,height=2.5in}
\caption{$B$ signal yields as a function of  
$\cos\theta_H$ for events in (a) $K^*(892)^+$, (b) $\rho(770)^-$,
(c) $K^{*0}_x$ and (d) $K^{*+}_x$ regions.  
Points are data and histograms are the expectations from (a) 
$B^0\to K^*(892)^+\pi^-$, (b) $B^0\to \rho(770)^- K^+$ and (c,d) $B$ decays to
a scalar and a pseudoscalar meson.
}
\label{fig:hel}
\end{center}
\end{figure}

To further understand the possible resonances, we study the 
distributions of $\cos\theta_H$, where the helicity angle $\theta_H$
is  defined as the angle between the direction of the candidate $B$ meson and
the $K^+$ ($\pi^0$) direction in the $K^*(\rho)$ rest frame. Figure 
\ref{fig:hel} shows the $B$ yields as a function of $\cos\theta_H$ for events
in the $K^*(892)^+, \rho(770)^-, K^*_x(K\pi)$ signal region. The 
$\cos\theta_H$ distributions for the first two modes are consistent with those
of $B^0 \to$  pseudoscalar vector (PV) decays, as expected for 
$B^0\to K^*(892)^+\pi^-$ and $B^0\to \rho(770)^- K^+$.
Note that the asymmetry in the  $\cos\theta_H$ distributions  is due to the
 inefficiency of low momentum $\pi^0$ reconstruction. Since $\pi^0$s from
$B^0\to \rho^- K^+$ decays are more energetic  than those from 
$B^0\to K^*(892)^+ \pi^-$, the asymmetric effect is less pronounced. The 
$\cos\theta_H$
distributions for the $K^{*0}_x$ and $K^{*+}_x$ modes 
 favor a scalar behavior.

\begin{figure}[t]
\begin{center}
\epsfig{file=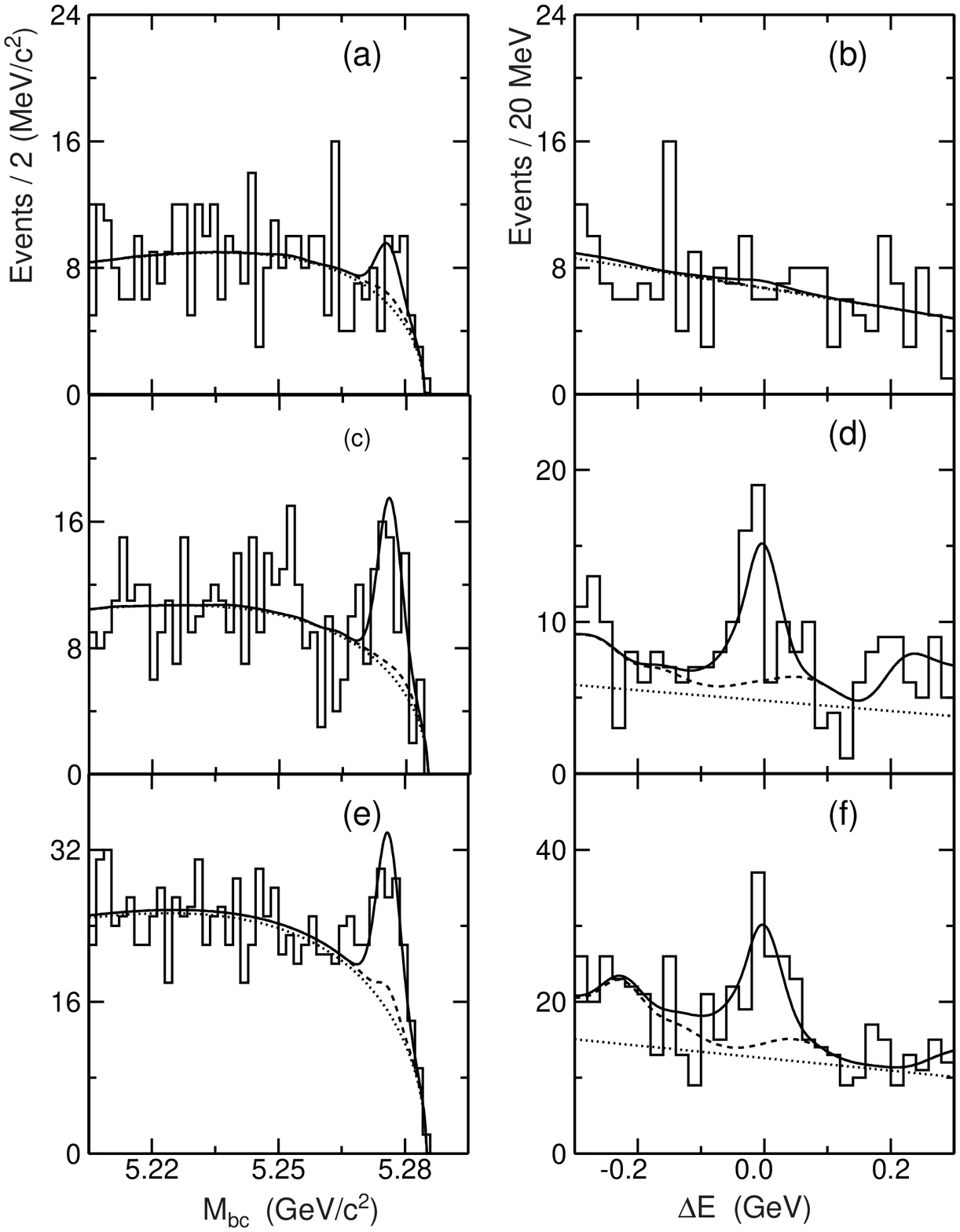,width=5in,height=5in}
\caption{$\Mbc$ and $\Delta E$ distributions for (a,b) $K^{*0}\pi^0,$ 
(c,d) $K^{*+}\pi^-$
and (e,f) $\rho^+ K^-$ events. The superimposed solid curves represent the
 corresponding fits. Dashed lines are the background projections and the dotted 
 lines show the continuum $q\bar{q}$ contribution. Enhancements in the projection curves
 at $\Delta E>0.2$ GeV in (d) and (f) are due to $B\to K\pi$ decays. }
\label{fig:mbde_sub}
\end{center}
\end{figure}

\section{Yield for Various States}

We measure $B^0\to K^+\pi^-\pi^0$ decay rates for the three PV modes,  
events in the two $K^*_x\pi$ regions, and the central 
region of the Dalitz plot with two-body masses above 2.0 GeV/$c^2$.   
Candidate $K^*(892)$ and $\rho(770)^-$ mesons are identified by requiring the
$K^+\pi^{-(0)}$ and $\pi^-\pi^0$ masses to be in the range 
820--980$\,\mathrm{MeV}/c^2$ and 570--970$\,\mathrm{MeV}/c^2$, respectively. 
To further reduce background, a selection of 
$|\cos\theta_H|>0.3$ is applied to the vector meson candidates.
In the two $K_x^{*0(+)}\pi$ regions, we require  
$1.1$ GeV/$c^2 < M(K^+\pi^{-(0)})<1.6$ GeV/$c^2$. $B$ meson candidates  are then
selected from the inclusive $K^+\pi^-\pi^0$ events after applying all analysis
cuts,
including the appropriate two-body mass vetos to avoid cross talk. 
The signal PDFs are obtained from MC 
simulations for all six channels, where a scalar hypothesis is used to model
$K^*_x \pi$.   

\begin{figure}[t]
\begin{center}
\epsfig{file=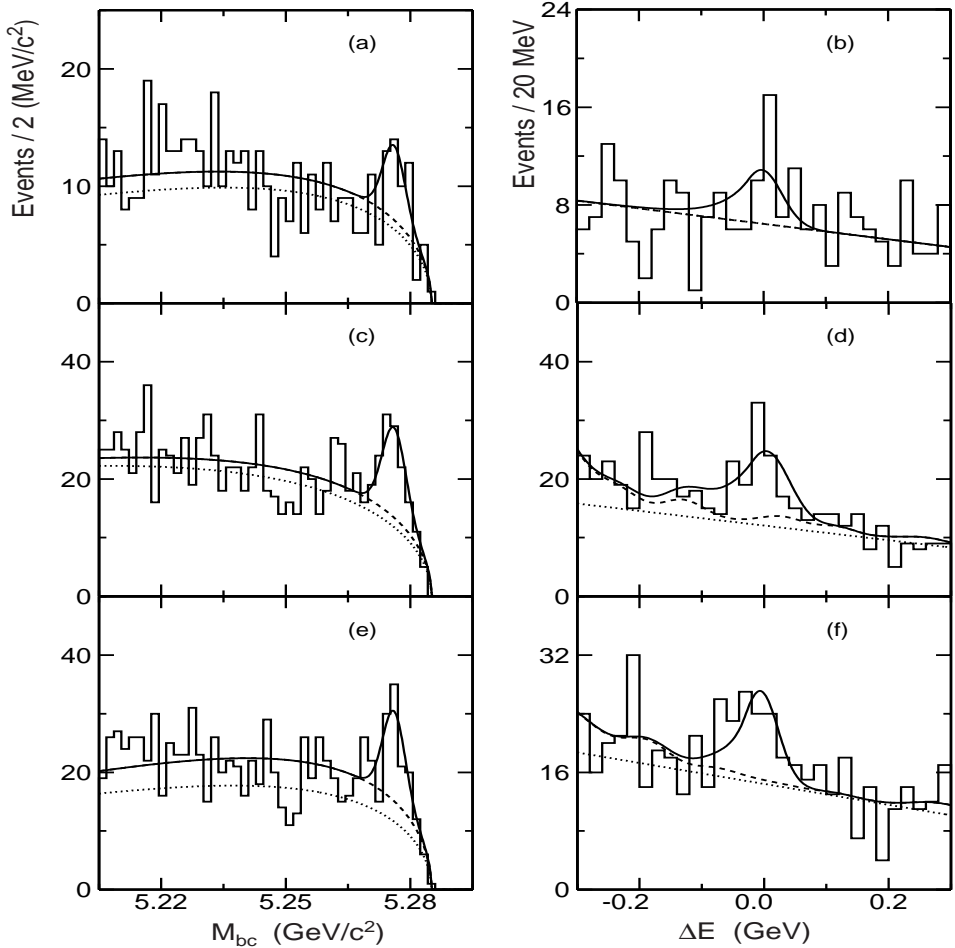,width=5in,height=5in}
\caption{$\Mbc$ and $\Delta E$ distributions for events in different regions
of the Dalitz plots: (a,b) the central region, (c,d) $K_x^{*0}\pi^0$ region
and  (e,f) $K_x^{*+}\pi^-$ region.
The superimposed solid curves represent the
 corresponding fits. Dashed lines are the background projections and the dotted 
 lines show the continuum $q\bar{q}$ contribution. Enhancements in the projection curves
 at $\Delta E>0.2$ GeV in (d) and (f) are due to $B\to K\pi$ decays }
\label{fig:mbde_sub2}
\end{center}
\end{figure}

Figure \ref{fig:mbde_sub} shows the $\Mbc$ and $\Delta E$ distributions,
and their corresponding fit curves, for the three PV modes. 
  No signal yield is seen in the
$K^{*0}\pi^0$ channel but
significant signals are observed for the $K^{*+}\pi^-$ and $\rho^- K^+$ 
modes; the yields measured from the $\Delta E$ fits are
  $38\pm 11$ and $77^{+18}_{-17}$ events, with  statistical
significances of $3.8 \sigma$ and $4.9 \sigma$, respectively. 
As for events in the central region of the Dalitz plot and the 
two $K^*_x$ regions,
Fig. \ref{fig:mbde_sub2} shows their $\Mbc$
and $\Delta E$ distributions  with the fit curves superimposed.  
Based on the $\Delta E$ fit, 
there are $67\pm 17$ and $52\pm 15$ signal events in 
 $K_x^{*0}\pi^0$ and $K_x^{*+}\pi^-$, respectively. Since events
in these $K_x^*\pi$ regions cannot be positively identified, their 
reconstruction efficiencies are determined without assumptions about the intermediate two-body states.  
Although a yield of
around 20
events is obtained from both the $\Delta E$ and $\Mbc$ fits for  the
central region of the Dalitz plot, the statistical significance is below 
$3\sigma$ and, hence, an upper limit is reported.  The reconstruction 
efficiency is obtained from a phase-space decay model.  
     
The number of feed-across events from high mass $K^*_x$ to  the $K^*(892)$  
region is estimated from the $B$ yields in the $1.1$ GeV/$c^2 < M(K^+\pi^0)<1.6$ GeV/$c^2$ region,
and the Breit-Wigner distribution, modeled with a mass of 1.326 GeV/$c^2$ and
a mean of 252 MeV/$c^2$. We find a contribution of 1 event. 
Assuming no  interference, the 
$K^*(892)^+\pi^-$ yield is estimated to be $37\pm 11$.
The possible effect of  interference is studied using a Monte Carlo simulation
which assumes the three $PV$ decays and two $K_x^{*}\pi$ states.
We compare the yields with and without interference.  After varying the relative phase of each channel,
and taking into account the non-uniform reconstruction efficiency over the 
Dalitz plot, the largest deviation is $^{+16\%}_{-6\%}$ for 
$K^*(892)^+\pi^-$ and $^{+13\%}_{-14\%}$ for $\rho^-K^+$. These two numbers
are used to estimate the systematic error arising from interference.

\begin{table}[t]
\begin{center}
\caption{Summary of the $B^0\to K^+\pi^-\pi^0$ search. We present signal yields,
efficiencies, and their statistical significances for the inclusive mode,
the three intermediate channels of $B\to $ PV decays and the other three regions
of the Dalitz plot. Branching fractions and/or upper limits are 
shown in the last two columns. In the branching fractions, the first 
error is statistical and the second systematic. The third error for 
$K^*(892)^+\pi^-$ and $\rho^- K^+$ corresponds to the largest uncertainty from
the interference between different states. The last channel, 
$K^+\pi^-\pi^0_{NR}$, indicates the  non-resonant $B^0\to K^+\pi^-\pi^0$ decay.   
}
\label{tbl:result}
\vspace{0.2 cm}
\begin{tabular}{|c|ccccc|} \hline
Channel & Yield  & Eff.(\%)& Sig.    & BF ($10^{-6}$) & UL ($10^{-6}$) \\ \hline
$K^+\pi^-\pi^0$ & $386\pm 44$ & 12.4 & 9.4 & 
$36.6^{+4.2}_{-4.1}\pm 3.0$ & - \\
$K^{*}(892)^0\pi^0$ & $2^{+11}_{-10}$ & 6.7&
0.3 & $0.4^{+1.9}_{-1.7}\pm 0.1$ & 3.5  \\
 $K^{*}(892)^+\pi^-$ & $37\pm 11$& 2.9& 3.8 &
   $14.8^{+4.6+1.5+2.4}_{-4.4-1.0-0.9}$ & -\\
$\rho^-K^+$ & $77^{+18}_{-17}$ & 6.0 & 4.9 &  $15.1^{+3.4+1.4+2.0}_{-3.3-1.5-2.1}$ & - \\
$K_x^{*0}\pi^0$ & $67\pm 17$ & 12.9 &4.2&  $6.1^{+1.6+0.5}_{-1.5-0.6}$ & - \\
$K_x^{*+}\pi^-$ & $52\pm 15$ & 11.9 &3.7& 
$5.1\pm 1.5^{+0.6}_{-0.7}$ & - \\
$K^+\pi^-\pi^0_{NR}$ & $22^{+10}_{-9}$ & 4.1 & 2.5 &  $5.7^{+2.7+0.5}_{-2.5-0.4}$  & $9.4$ \\\hline
\end{tabular}
\end{center}
\end{table}
       
\section{Systematic Uncertainties}

The systematic error for each signal yield is estimated 
by varying each parameter of the fit functions 
by $\pm 1 \sigma$ from the measured values.
The shifts in signal yield are then added in quadrature.
The typical fit systematic error for the inclusive decay is around 4\%.
Signal efficiencies are first obtained from MC simulations,
and then corrected by comparing data and MC predictions for other processes.
The efficiency for the inclusive $K^+\pi^-\pi^0$ signals is estimated from
the weighted  sum of the efficiencies for the possible two-body intermediate states shown in
Fig. \ref{fig:deind}, where the sub-decay branching fraction for each 
two-body state is not included. The uncertainty on this inclusive efficiency is
5\%, determined by checking the reconstruction efficiencies on various 
two-body modes.       
The $\pi^0$ reconstruction efficiency is verified by comparing the $\pi^0$
decay angular distribution with the MC prediction, and by
measuring the ratio of 
the branching fractions of two $\eta$ decay channels: $\eta\to \gamma\gamma$ 
and $\eta\to \pi^0\pi^0\pi^0$. The typical systematic error for $\pi^0$ 
detection is 3\%.
The systematic errors on the charged track reconstruction  
are estimated to be $\sim 2$\% using  partially 
reconstructed $D^*$ events, and  verified by comparing the ratio of 
$\eta\to \pi^+\pi^-\pi^0$ to $\eta\to \gamma\gamma$ 
in data with MC expectations. The final systematic errors on the reconstruction
efficiencies, including charged particle and $\pi^0$ detection, particle 
identification  and the $\LR$ cut,  range from  
7 to 14\% for the quasi-two-body decays and is 8.2\% for the inclusive 
$K^+\pi^-\pi^0$ channel.

\section{Branching Fractions}

Table \ref{tbl:result} summarizes 
the fit results for each reconstructed decay channel. The branching fractions 
and upper limits are calculated assuming that $B^+B^-$ and 
$B^0\overline{B}{}^0$ are produced with equal probability.
The systematic errors on the branching fractions combine
the systematic errors for the  reconstruction efficiencies and the $\Delta E$ fit
with the uncertainty from the number of $B\overline{B}$ events. 
Since no signal is seen in the $K^{*0}\pi^0$ mode, and 
the signal yield in the central region of the Dalitz plot is not significant,
upper limits are computed at the 90\% confidence level (C.L.) based on
the observed number of events in the signal region, and the background level 
found by the fit; both the statistical and systematic errors are taken into
account~\cite{limit-ref}.
With 85.0 million $B\overline{B}$ events, 
we measure the branching fraction to be
$(36.6^{+4.2}_{-4.1}({\rm stat.})\pm 3.0({\rm syst.}))\times 10^{-6}$
 for the inclusive $B^0\to K^+\pi^-\pi^0$ decay, without assumptions about 
intermediate two-body states.    
The branching fractions of $B^0\to K^*(892)^+\pi^-$ and $B^0\to \rho^- K^+$
decay are measured to be $(14.8^{+4.6+1.5+2.4}_{-4.4-1.0-0.9})\times 10^{-6}$
 and
$(15.1^{+3.4+1.4+2.0}_{-3.3-1.5-2.1})\times 10^{-6}$,
where the first error is statistical, the second is systematic and the third
corresponds to the largest uncertainty from the  interference between
different states. Finally, the $B^0\to K^+\pi^-\pi^0$ decay branching fractions
 in the 
$K_x^{*0}\pi^0$ and $K_x^{*+}\pi^-$ regions are measured to be 
$(6.1^{+1.6+0.5}_{-1.5-0.6})
\times 10^{-6}$ and $(5.1\pm 1.5 ^{+0.6}_{-0.7})\times 10^{-6}$, respectively.  

\begin{figure}[t]
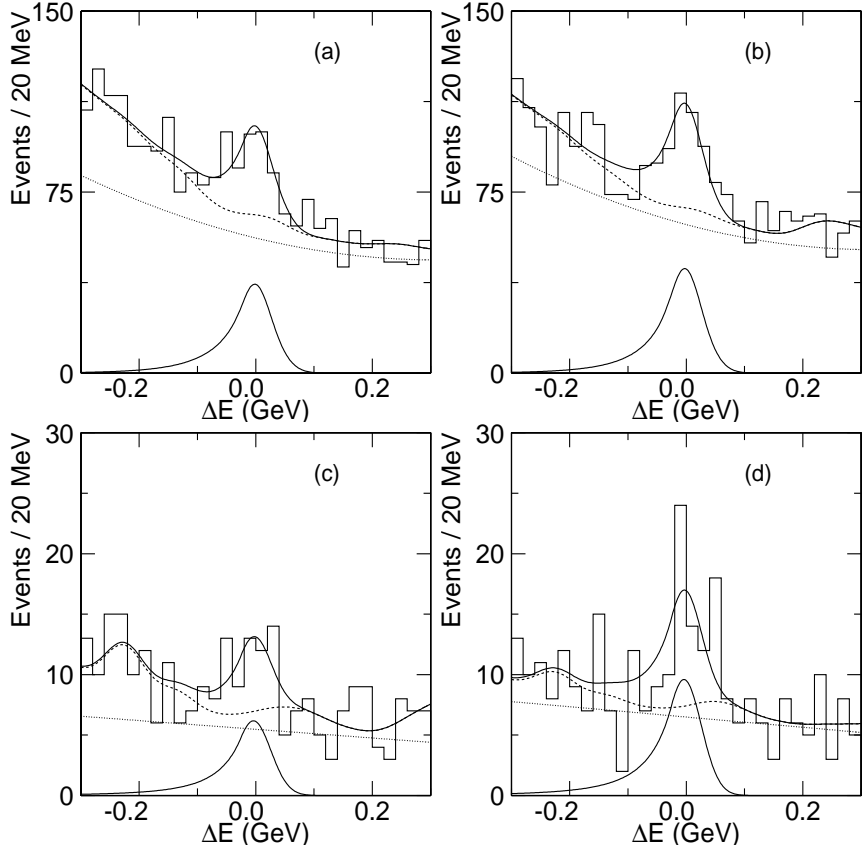

\begin{center}
\epsfig{file=de_acp_p_inc.epsi,width=2.2in,height=2.2in}
\epsfig{file=de_acp_m_inc.epsi,width=2.2in,height=2.2in}
\epsfig{file=de_acp_p_rhok.epsi,width=2.2in,height=2.2in}
\epsfig{file=de_acp_m_rhok.epsi,width=2.2in,height=2.2in}
\caption{ $\Delta E$ distributions for the (a) $K^+\pi^-\pi^0$, (b) 
$K^-\pi^+\pi^0$, (c) $\rho^- K^+$ and (d) $\rho^+K^-$ samples. 
The superimposed solid curves represent the
 corresponding fits and signal projections. 
Dashed lines are the background projections and dotted 
 lines show the continuum $q\bar{q}$ contribution. Enhancements in the projection curves
 at $\Delta E>0.2$ GeV  are due to $B\to K\pi$ decays }
\label{fig:acp}
\end{center}
\end{figure}

\section{Search for $CP$ Violation}

Using the large signals observed in the inclusive $B^0\to K^+\pi^-\pi^0$
and $B^0\to \rho^- K^+$ modes,
we search for  direct $CP$ violation  by dividing the data into two
subsets, according to the charge of the kaon. The asymmetry, defined as
$A_{CP} = \frac{N_{\bar{B}} -N_B}{N_{\bar{B}} +N_B}$, 
is then computed using the $B$ 
signal yields obtained  from $\Delta E$ fits. Following the same fitting 
procedure, we observe 
$179^{+31}_{-30}\; K^+\pi^-\pi^0$ events and 
$207^{+32}_{-31}\; K^-\pi^+\pi^0$ events, while  
the $\Delta E$ fit yields $30^{+12}_{-11}$ $\rho^- K^+$ events
and $47^{+13}_{-12}$ $\rho^+ K^-$ events (see Fig.\ref{fig:acp}). 
The possible reconstruction
bias in $A_{CP}$ is studied by checking the inclusive
 $\overline{D}{}^0\to K^+\pi^-$ and $D^0\to K^-\pi^+$ yields in data. The 
obtained systematic error is $0.5\%$. Adding this $0.5\%$ error in quadrature 
with
the fitting systematic error, obtained by varying each parameter in the PDFs 
by  $1 \sigma$, gives the total systematic error. 
Finally, the $CP$ asymmetry is calculated to be $A_{CP}=0.07\pm 0.11 \pm 0.01$
for the inclusive mode, and $A_{CP}=0.22^{+0.22+0.06}_{-0.23-0.02}$ for
$B^0 (\overline{B}{}^0)  \to \rho^\mp K^\pm$. We also set 90\% confidence intervals on the  asymmetry of
$-0.12<A_{CP}<0.26$ for the inclusive mode, and $-0.18<A_{CP}<0.64$ for 
$B^0\to\rho^\mp K^\pm$.

\section{Conclusions}

In summary, we have studied the charmless hadronic  decays,
$B^0\to K^+ \pi^-\pi^0$, which is observed for the first time. 
Our results show that the branching fraction of $B^0\to K^+\pi^-\pi^0$ is  
$(64\pm 10)\%$ and $(78\pm 15)\%$ of that of 
$B^+\to K^+\pi^-\pi^+$ \cite{kshh,bkhh} and $B^0\to K^0\pi^-\pi^+$ 
\cite{kshh,ckstp}, respectively. 
The $K^+\pi^-\pi^0$ signal candidates populate
 the edge of the Dalitz plot, indicating the existence of 
quasi-two-body states. For the $K^+\pi^-\pi^0$ final state, 
we have observed  signals in the $K^{*+}\pi^-$ and $\rho^-K^+$ samples 
but no significant $K^{*0}\pi^0$  signal is seen. The $\rho^- K^+$ branching 
fraction is close to the $K^{*+}\pi^-$
branching fraction, where our measurement is consistent with the earlier 
CLEO result \cite{ckstp}. However, our $\rho^- K^+$ result is twice that of
BaBar's measurement \cite{brhok}.  
We also  report the $B$ decay rates
in other  regions of the Dalitz plot without assumptions about  the presence of
two-body intermediate states. In the future, significantly more data will be 
collected at Belle, which
will enable us to perform a full Dalitz analysis, allowing
us to identify other quasi-two-body states and extract their 
relative phases. Finally,  we performed a search for direct CP violation in
the inclusive and $B^0\to \rho^- K^+$ channels. 
No evidence of $CP$ violating asymmetry is seen and  
90\% C.L. limits on $A_{CP}$ are set. 

{\bf Acknowledgements} 

We wish to thank the KEKB accelerator group for the excellent
operation of the KEKB accelerator.
We acknowledge support from the Ministry of Education,
Culture, Sports, Science, and Technology of Japan
and the Japan Society for the Promotion of Science;
the Australian Research Council
and the Australian Department of Education, Science and Training;
the National Science Foundation of China under contract No.~10175071;
the Department of Science and Technology of India;
the BK21 program of the Ministry of Education of Korea
and the CHEP SRC program of the Korea Science and Engineering Foundation;
the Polish State Committee for Scientific Research
under contract No.~2P03B 01324;
the Ministry of Science and Technology of the Russian Federation;
the Ministry of Education, Science and Sport of the Republic of Slovenia;
the National Science Council and the Ministry of Education of Taiwan;
and the U.S.\ Department of Energy.

\newpage

\end{document}